\documentclass[twocolumn,pra,showpacs,amsmath,amssymb]{revtex4-2}
\usepackage{graphicx}
\usepackage{bm}
\usepackage{epstopdf}
\usepackage{epsfig}
\usepackage{dsfont}
\usepackage{amssymb}
\usepackage{siunitx}
\usepackage{tikz}
\usetikzlibrary{positioning}
\usetikzlibrary{shapes}
\usepackage{diagbox}
\newcommand{\scale}{0.9}
\usepackage{layout}

\begin{document}

\title{Numerically assisted determination of local models in network scenarios}
\author{José Mário da Silva}
\email[]{jose.filho@ufpe.br}
\author{Fernando Parisio}
\email[]{fernando.parisio@ufpe.br}
\affiliation{Departamento de
F\'{\i}sica, Universidade Federal de Pernambuco, Recife, Pernambuco
50670-901 Brazil}

\begin{abstract}
Taking advantage of the fact that the cardinalities of hidden variables in network scenarios can be assumed to be finite without loss of generality, a numerical tool for finding explicit local models that reproduce a given statistical behavior was developed. The numerical procedure was then validated using families of statistical behaviors for which the network-local boundary is known, in the bilocal scenario. Furthermore, the critical visibility for three notable distributions mixed with a uniform random noise is investigated in the triangle network without inputs. We provide conjectures for the critical visibilities of the Greenberger–Horne–Zeilinger and W distributions (which are roots of fourth degree polynomials), as well as a lower bound estimate of the critical visibility of the elegant joint measurement distribution. The developed codes and documentation are publicly available at \href{https://github.com/mariofilho281/localmodels/tree/master}{\textcolor{blue}{github.com/mariofilho281/localmodels}}.
\end{abstract}
\maketitle

\section{Introduction}
\label{intro}
Nonlocality is a hallmark feature of quantum theory that differentiates it from any classical conceptual framework \cite{Bell}. Soon after Bell's seminal discovery in 1964, much attention was given to nonlocality in the simplest standard Bell scenario, i.e., a situation where a single source distributes information (classical or quantum) to two parties. Over the years, multipartite scenarios, where one source distributes information to several parties, have also been studied \cite{multipartite1,multipartite2,multipartite3,multipartite4}. Not only has nonlocality had a profound impact in the foundations of physics, but it also has found application in the field of quantum information \cite{Ekert91,nonlocality_and_communication_complexity,revision_bell,scarani}.

The next conceptual leap occurred in 2010 with the introduction of nonlocality in network scenarios \cite{bilocality}. In these scenarios there are several sources, assumed to be independent, distributing information to some of the parties (not all) forming a network topology. All examples of quantum nonlocality in the standard Bell scenarios have a direct analog in network scenarios \cite{correlation_scenarios}, but some cases of network-nonlocality do not seem to stem directly from the violation of a Bell inequality in a standard scenario. A fascinating illustration is the occurrence of nonlocality in situations where all parties have no input choices \cite{RGB4}. This may indicate that standard Bell nonlocality is just a specific subset of the more general phenomenon of network-nonlocality, which is far from being completely understood \cite{revision_networks}.

An important step towards a better understanding of network nonlocality was the demonstration that network-local correlations, with a finite number of inputs and outputs, can always be represented by models where the sources distribute hidden variables with finite cardinality, i.e., variables that belong to a finite set of possibilities \cite{cardinality_bound}. A pivotal problem in the field is to know whether a given statistical behavior is network-local or not. If it is, the referred result guarantees that it can be reproduced by a model where the hidden variables only take on a finite number of values. Such models can be represented by finite arrays of real numbers, and therefore can be subjected to optimization algorithms that aim at fitting a local model to the given behavior. 

The purpose of this paper is to present an implementation of this idea, as well as the results we obtained when applying the procedure to certain statistical behaviors in the bilocal and triangle network topologies. In these relatively simple scenarios we were able to devise several closed analytical models from the numerical results. The behaviors we consider are not limited to quantum mechanical ones, but may refer to distributions achievable with supraquantum resources.

The structure of this paper is as follows: in Sec. \ref{sec:local models}, the representation of local models with finite hidden variable cardinalities is explained, as well as the general procedure to obtain models for a given statistical behavior; in Sec. \ref{sec:bilocal scenario numerical}, the procedure is applied to statistical behaviors in the bilocal scenario; in Sec. \ref{sec:triangle scenario numerical}, the procedure is applied to noteworthy probability distributions in the triangle scenario; in Sec. \ref{closingR} we make our final remarks and discuss some possibilities for future work.

\section{Representation of local models}
\label{sec:local models}

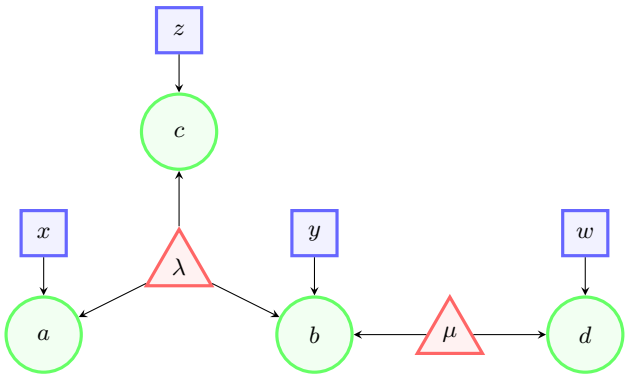
\begin{figure}
\centering
\begin{tikzpicture}[scale=\scale,
		output/.style={circle, draw=green!60, fill=green!5, very thick, minimum size=10mm},
		hidden/.style={regular polygon, regular polygon sides=3, draw=red!60, fill=red!5, very thick, inner sep=0,minimum size=10mm},
		input/.style={rectangle,draw=blue!60, fill=blue!5, very thick, minimum size=6mm}
		]
	%Nodes
	\node[output] (a) at (0,0) {$a$};
	\node[input] (x) at (0,1.5) {$x$};
	\node[hidden] (lambda) at (2,1) {$\lambda$};
	\node[output] (b) at (4,0) {$b$};
	\node[output] (c) at (2,3) {$c$};
	\node[input] (z) at (2,4.5) {$z$};
	\node[hidden] (mu) at (6,0) {$\mu$};
	\node[output] (d) at (8,0) {$d$};
	\node[input] (y) at (4,1.5) {$y$};
	\node[input] (w) at (8,1.5) {$w$};
	
	%Edges
	\draw[-stealth] (x.south) -- (a.north);
	\draw[-stealth] (lambda) -- (a);
	\draw[-stealth] (lambda) -- (b);
	\draw[-stealth] (lambda) -- (c);
	\draw[-stealth] (z) -- (c);
	\draw[-stealth] (mu.west) -- (b.east);
	\draw[-stealth] (mu.east) -- (d.west);
	\draw[-stealth] (y.south) -- (b.north);
	\draw[-stealth] (w) -- (d);
\end{tikzpicture}
\caption{Example of a network scenario. The outputs of each party, denoted by $a$, $b$, $c$, and $d$, are represented by green circles, while the inputs $x$, $y$, $z$, and $w$ are represented by blue squares. The local hidden variables $\lambda$ and $\mu$ are represented by red triangles. A network-local behavior $p(a,b,c,d|x,y,z,w)$ in such a scenario can always be cast into the form of Eq. (\ref{network local}) (notice the statistical independence between $\lambda$ and $\mu$).}
\label{fig: rede}
\end{figure}

Network scenarios are situations where there are multiple independent sources distributing information to several parties. We start this section by illustrating the distinctive features of these scenarios with a specific example with four parties and two independent sources, see Fig. \ref{fig: rede}. Alice and Charles receive information from one of the sources, whereas Dave gets information from the other source. In this example, the only party that receives information from both sources is Bob. In general, all parties have measurement settings, also called inputs. The outputs $a$, $b$, $c$, $d$ have finite cardinalities that will be represented by $m_a$, $m_b$, $m_c$, $m_d$, and the inputs $x$, $y$, $z$, $w$ also have finite cardinalities, represented by $M_a$, $M_b$, $M_c$, $M_d$. Throughout the paper, whenever a variable $\lambda$ has finite cardinality $c_{\lambda}$, it will be assumed to take on values in the set $\{0,1,\ldots,c_{\lambda}-1\}$. The statistical behavior $p(a,b,c,d|x,y,z,w)$ is said to be network-local if and only if there are probability distributions $p_\lambda(\lambda)$, $p_\mu(\mu)$, $p_a(a|x,\lambda)$, $p_b(b|y,\lambda, \mu)$, $p_c(c|z,\lambda)$ and $p_d(d|w,\mu)$, such that
\begin{multline}
    p(a,b,c,d|x,y,z,w) = \int d\lambda\, d\mu\,p_\lambda(\lambda) p_\mu(\mu)\\
    p_a(a|x,\lambda) p_b(b|y,\lambda, \mu) p_c(c|z,\lambda) p_d(d|w,\mu).\label{network local}
\end{multline}
There are other stronger notions of network nonlocality such as requiring that all sources distribute nonlocal resources \cite{full}, but in this paper we work with the more general definition of network nonlocal behaviors as simply the ones that cannot be cast in the form of Eq.  (\ref{network local}).

As mentioned before, in Ref. \cite{cardinality_bound} it was proven that the cardinalities of local hidden variables in networks can always be assumed to be finite. The authors also provide an upper bound for the necessary cardinalities that we reproduce here for completeness. Consider for example the hidden variable $\lambda$ of Fig. \ref{fig: rede}. We begin by identifying the full probability distribution of the network $p(a,b,c,d|x,y,z,w)$ and the probability distribution of the parties not connected to the hidden variable, $p(d|w)$, which in this case is only Dave's. The dimensions of these probability distributions are then calculated using the Collins-Gisin parametrization \cite{Collins-Gisin}. This is basically the number of entries in each distribution minus the number of constraints imposed by no-signalling and normalization conditions \cite{cardinality_bound}:
\begin{align*}
	\dim\left\{p(a,b,c,d|x,y,z,w)\right\} &= \prod_{i=a,b,c,d}[M_i(m_i-1)+1]-1,\\
	\dim\left\{p(d|w)\right\} &= M_d(m_d-1).
\end{align*}
Finally, the cardinality upper bound is given by the difference of these two dimensions:
\begin{align}
    c_\lambda &\leq \dim\left\{p(a,b,c,d|x,y,z,w)\right\} - \dim\left\{p(d|w)\right\}\label{cardinality bound}.
\end{align}
Of course we can always count the deterministic strategies available to parties that only receive $\lambda$ (Alice and Charles in this example) and bound $c_\lambda$ even further (similar to what is done in the discussion about the local polytope in Ref. \cite{revision_bell}):
\begin{align}
    c_\lambda &\leq m_a^{M_a},\label{cardinality bound Alice det}\\
    c_\lambda &\leq m_c^{M_c}.\label{cardinality bound Charles det}
\end{align}
The lowest number in inequalities (\ref{cardinality bound}), (\ref{cardinality bound Alice det}), and (\ref{cardinality bound Charles det}) could then be considered the final upper bound on $c_\lambda$.

If the hidden variables $\lambda$ and $\mu$ have finite cardinalities $c_\lambda$ and $c_\mu$, Eq. (\ref{network local}) reduces to
\begin{multline}
    p(a,b,c,d|x,y,z,w) = \sum_{\lambda=0}^{c_\lambda-1}\sum_{\mu=0}^{c_\mu-1} p_\lambda(\lambda) p_\mu(\mu)\\
    p_a(a|x,\lambda) p_b(b|y,\lambda, \mu) p_c(c|z,\lambda) p_d(d|w,\mu),\label{network local finite}
\end{multline}
and the model can be represented by finite arrays of numbers representing the hidden-variable distributions $p_\lambda$, $p_\mu$ and the parties response functions $p_a$, $p_b$, $p_c$, and $p_d$. To avoid notation clutter, the arrays will be represented by the same symbols as the distributions, but indexed with brackets, like $p_\lambda[i]$, for example. The hidden variable arrays are just lists of the probabilities that the variable assumes the index value:
\begin{equation}
p_\lambda[i] = P(\lambda = i).\label{hidden variable array}
\end{equation}

For example, a hidden variable characterized by $p_\lambda = \begin{bmatrix}3/4 & 1/4\end{bmatrix}$ is one that produces 0 with probability 3/4 or 1 with probability 1/4. Obviously, the last element of the array is not independent of the others because of the normalization condition $\sum_{i=0}^{c_\lambda-1}p_\lambda[i]=1$. Because of this, only $c_\lambda - 1$ probabilities need to be subjected to the optimization procedure when trying to fit a model to a statistical behavior, as long as we enforce the inequality constraint $\sum_{i=0}^{c_\lambda-2}p_\lambda[i] \leq 1$ in order to have the last element be a valid probability. Although the optimization procedure only works with $c_\lambda -1$ elements for a given $p_\lambda$, we will provide the full array when reporting specific models in Secs. \ref{sec:bilocal scenario numerical} and \ref{sec:triangle scenario numerical}.

The response function arrays need to have more indices, like $p_b[i,j,k,l]$, for the example of Fig. \ref{fig: rede}. The convention we adopt is as follows: the output value as the first index, the input value as the second index (if the party has a choice of measurement settings), and the remaining indices being the values of the hidden variables connected to that party. 

The generalization of this procedure to arbitrary network architectures is immediate. The response function array of a generic party with output $a$, input $x$ and hidden variables $\lambda_1,\ldots,\lambda_n$ is then defined as:
\begin{equation}
    p_a[i,j,k,\ldots,l] = P(a=i|x=j,\lambda_1=k,\ldots,\lambda_n = l).\label{response function array}
\end{equation}

Because of the normalization condition in the response function $\sum_{i=0}^{m_a-1}p_a[i,j,k,\ldots,l]=1$, the numerical procedure needs to optimize only $m_a-1$ probabilities for each combination of inputs and hidden variables. In summary, the number of parameters to optimize are $c_\lambda-1$ per hidden variable and $(m_a-1)M_ac_{\lambda_1}\cdots c_{\lambda_n}$ per party.

When reporting response functions of parties, we will do so in matrix form, using the second to last index to represent rows and the last index to represent columns. In our example, if Alice's output $a$ is binary, the response function array
\[p_a[0,x,\lambda] = \begin{bmatrix}
	  1 & \frac{1}{2}\\
	\frac{1}{2} & 0
\end{bmatrix}\]
represents a situation where Alice outputs $x$ if $x=\lambda$ or a uniformly distributed random bit if $x\neq \lambda$. The array $p_a[1,x,\lambda]$, being determined by the normalization condition.

Given the hidden-variable arrays $p_\lambda,p_\mu,\ldots$ and the party arrays $p_a,p_b,\ldots$, all the probabilities $p(a,b,\ldots|x,y,\ldots)$ can be calculated:
\begin{multline}
p(a,b,\ldots|x,y,\ldots) = \\\sum_{i,j,\ldots}p_\lambda[i]p_\mu[j]\ldots p_a[a,x,\ldots]p_b[b,y,\ldots]\ldots.\label{local behavior with finite hidden variable cardinalities}
\end{multline}
This is just the generalization of Eq. (\ref{network local finite}) where no specific network topology is assumed.

Therefore, the problem of finding a local model for a given behavior $p^*(a,b,\ldots|x,y,\ldots)$ can be reduced to a constrained non-linear optimization problem on the elements of $p_\lambda,p_\mu,\ldots$ and $p_a,p_b,\ldots$, with the sum of squared errors in the probabilities as the cost function:
\begin{equation}
\begin{aligned}
\min_{\substack{p_\lambda,p_\mu,\ldots,\\p_a,p_b,\ldots}} &\sum_{\substack{a,b,\ldots,\\x,y,\ldots}} \left[p(a,b,\ldots|x,y,\ldots) - p^*(a,b,\ldots|x,y,\ldots)\right]^2\\
\text{s.t.}\qquad &p_\lambda,p_\mu,\ldots,p_a,p_b,\ldots \geq 0\\
&\sum_i p_\lambda[i]\leq 1\\
&\sum_i p_\mu[i]\leq 1\\
&\cdots\\
&\sum_i p_a[i,\ldots] \leq 1\\
&\sum_i p_b[i,\ldots] \leq 1\\
&\cdots\\
\text{where }& p(a,b,\ldots|x,y,\ldots) \text{ are given by Eq. (\ref{local behavior with finite hidden variable cardinalities})}.
\end{aligned}\label{optimization problem}
\end{equation}

The sum of squared errors was chosen as the cost function mainly because of ease of numerical computation. However, it is possible to use other more common statistical distances without any relevant changes in the obtained results. For example, we have been able to reproduce the main results of Sec. \ref{subsec: W} using the Hellinger distance \cite{hellinger}.

In order to apply this idea to solve interesting problems regarding the bilocal and triangle topologies (Figs. \ref{fig: bilocal network} and \ref{fig:triangle scenario}), two \textsc{python} modules named bilocal.py and triangle.py were created. The code and documentation are publicly available at \href{https://github.com/mariofilho281/localmodels/tree/master}{\textcolor{blue}{github.com/mariofilho281/localmodels}}. In these two modules, the optimization problem \ref{optimization problem} is solved with a trust-region algorithm for constrained optimization available in the \textsc{scipy} package \cite{SciPy}. This algorithm deals with inequality constraints by employing the barrier method, which reduces the original problem to a sequence of equality constrained problems. Each of these subproblems is then solved using a trust-region sequential quadratic programming (SQP) with the projected conjugate gradient (CG) method \cite{trust-constr}.

One important remark is that in standard Bell nonlocality, the problem of finding a local model for a given behavior (if one such model exists) is substantially easier. In fact, it reduces to a feasibility problem in linear programming, for which there are algorithms that can provide an explicit model in case the behavior is local, or a Bell inequality violated by the behavior in case it is not local \cite{scarani}. In network nonlocality, such methods are not applicable because the problem of finding a network-local model is no longer an instance of convex optimization.

The idea of this work is in the same spirit of Ref. \cite{neural_network}, where the authors propose a machine learning algorithm to find local models for target distributions. These local models have hidden variables with continuous uniform distribution, and all the complexities needed to fit a given behavior falls onto the parties response functions, given by neural networks. In contrast, our method employs discrete hidden variables (in general not uniformly distributed), which allowed us to find simpler models, some even with closed analytical expressions for the probabilities involved (see Sec. \ref{sec:triangle scenario numerical}).

Finally, we would like to compare our method to that outlined in Ref. \cite{bilocality_extended}, Sec. II C, where the authors propose to optimize over correlators instead of probabilities, a methodology commonly used by researchers working in the context of the bilocal scenario. This strategy needs fewer optimization parameters than ours, so we expect faster execution times when optimizing over correlators instead of probabilities. However, it relies on the marginal independence of two parties' outcomes. We wanted to create a numerical tool that could be applied with little code modification to any network topology, including ones where such marginal independence is not present, e.g., the triangle scenario. Because of that, we have chosen to optimize over probabilities instead of correlators, trading computational efficiency for range of applicability.

\section{Bilocal network}
\label{sec:bilocal scenario numerical}

\begin{figure}
\centering
\begin{tikzpicture}[scale=\scale,
		output/.style={circle, draw=green!60, fill=green!5, very thick, minimum size=10mm},
		hidden/.style={regular polygon, regular polygon sides=3, draw=red!60, fill=red!5, very thick, inner sep=0,minimum size=10mm},
		input/.style={rectangle,draw=blue!60, fill=blue!5, very thick, minimum size=6mm}
		]
	%Nodes
	\node[output] (a) at (0,0) {$a$};
	\node[input] (x) at (0,1.5) {$x$};
	\node[hidden] (lambda) at (2,0) {$\lambda$};
	\node[output] (b) at (4,0) {$b$};
	\node[output] (c) at (8,0) {$c$};
	\node[input] (z) at (8,1.5) {$z$};
	\node[hidden] (mu) at (6,0) {$\mu$};
	\node[input] (y) at (4,1.5) {$y$};
	
	%Edges
	\draw[-stealth] (x.south) -- (a.north);
	\draw[-stealth] (lambda) -- (a);
	\draw[-stealth] (lambda) -- (b);
	\draw[-stealth] (z) -- (c);
	\draw[-stealth] (mu.west) -- (b.east);
	\draw[-stealth] (mu.east) -- (c.west);
	\draw[-stealth] (y.south) -- (b.north);
\end{tikzpicture}
\caption{In the bilocal scenario there are three parties and two sources. Alice (Charles) receives information from source $\lambda$ ($\mu$). Bob is connected to both sources. In contrast to standard Bell scenarios, the multiple hidden variables are not distributed to all parties and are assumed to be statistically independent.}
\label{fig: bilocal network}
\end{figure}
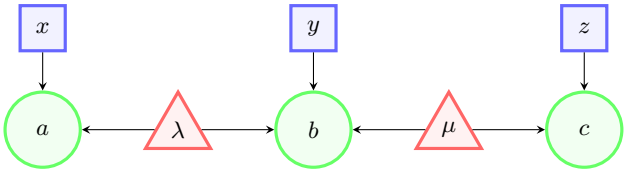

As a first application of the technique, let us consider the problem of finding an approximation for the boundary of the bilocal set in affine subspaces of dimension 2 in the no-signalling space. Following Ref. \cite{bilocality_extended}, consider the bilocal network (Fig. \ref{fig: bilocal network}) with binary inputs and outputs, and define the distributions $p_I$, $p_J$, and $p_0$ given by:

\begin{align}
    p_I(a,b,c|x,y,z) &= \frac{1}{8}\left[1 + \delta_{y,0}(-1)^{a+b+c}\right],\label{pI}\\
    p_J(a,b,c|x,y,z) &= \frac{1}{8}\left[1 + \delta_{y,1}(-1)^{x+z+a+b+c}\right],\label{pJ}\\
    p_0(a,b,c|x,y,z) &= \frac{1}{8}.\label{p0}
\end{align}

The behavior composed by a convex sum of the previous distributions $p = Ip_I + Jp_J + (1-I-J)p_0$ (with $|I|\le 1$ and $|J|\le 1$) is network local if and only if the so called BRGP inequality \cite{bilocality_extended} is satisfied:
\begin{equation}
    \sqrt{|I|} + \sqrt{|J|} \leq 1\label{BRGP}.
\end{equation}

\begin{figure}
	\centering
\includegraphics[width=\columnwidth]{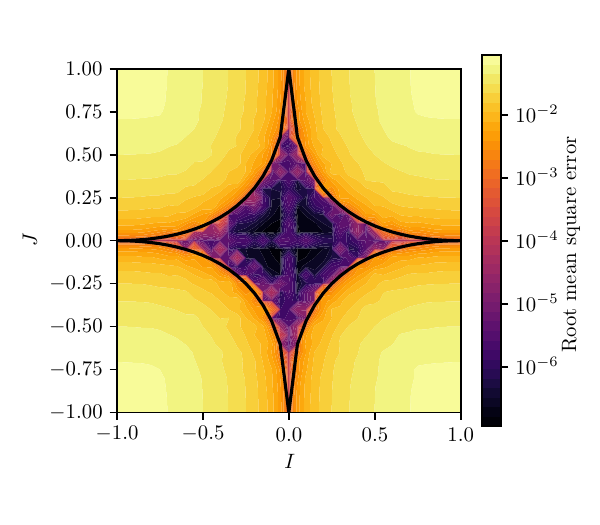}
\caption{Optimization error in the affine subspace spanned by the distributions (\ref{pI})--(\ref{p0}). The solid black line is the bilocal boundary given by the BRGP inequality (\ref{BRGP}). For behaviors inside the boundary, the solver can get the optimization error close to zero. As we move away from the boundary to behaviors that are not bilocal, the error grows, as expected. The color scale is logarithmic to convey information about the error values inside the bilocal boundary (in a linear scale, this region would display greater contrast as compared to the non-bilocal region).}
\label{fig:Bilocal error}
\end{figure}

By counting the deterministic strategies available to Alice and Charles in the bilocal scenario with two possible outputs, we can show that any network-local distribution can be achieved with the hidden variables cardinalities $c_\lambda=c_\mu=4$ \cite{cardinality_bound}.  We solved problem \ref{optimization problem} with these cardinalities for a rectangular grid of pairs $(I,J)$. The optimization error can be seen in Fig. \ref{fig:Bilocal error}. The optimization error of problem (\ref{optimization problem}) is first divided by 64 [the number of entries in the probability distribution $p(a,b,c|x,y,z)$] and then taken the square root, before being plotted. One can therefore interpret the color scale of Fig. \ref{fig:Bilocal error} as the root-mean-square error in the probabilities. The solid black line is the boundary of the network-local set in this two-dimensional slice given by inequality (\ref{BRGP}). The first thing we notice is that the points inside this region exhibit optimization error close to zero, showing that the solver has found acceptable local models for them. Also, behaviors outside the network-local boundary exhibit higher error, showing that the solver did not find faithful local models for them. This provides a first validation of the method in a known problem. 

Now, let us move on to a distinct two-dimensional slice of the behavior space. Let us define the following conditional distributions:
\begin{align}
p_X(a,b,c|x,y,z) &= \frac{1}{8}\left(\frac{1}{2}+\delta_{a,0}\right)\left[1 + \delta_{y,0}(-1)^{a+b+c}\right],\label{pX}\\
p_Y(a,b,c|x,y,z) &= \frac{1}{8}\left(\frac{1}{2}+\delta_{a,0}\right)\left[1 + \delta_{y,1}(-1)^{z+a+b+c}\right],\label{pY}\\
p_0(a,b,c|x,y,z) &= \frac{1}{8}\left(\frac{1}{2}+\delta_{a,0}\right).\label{p0 nontight}
\end{align}

For the remainder of this section, we will concern ourselves with the affine subspace spanned by the distributions (\ref{pX})--(\ref{p0 nontight}), i.e., the set of behaviors $p(a,b,c|x,y,z)$ given by the convex combination

\begin{equation}
p = Xp_X + Yp_Y + (1-X-Y)p_0,\label{subspace 2}
\end{equation}
where the dependency on $a$, $b$, $c$, $x$, $y$ and $z$ has been omitted to keep the notation clean.

\begin{figure}
	\centering
\includegraphics[width=\columnwidth]{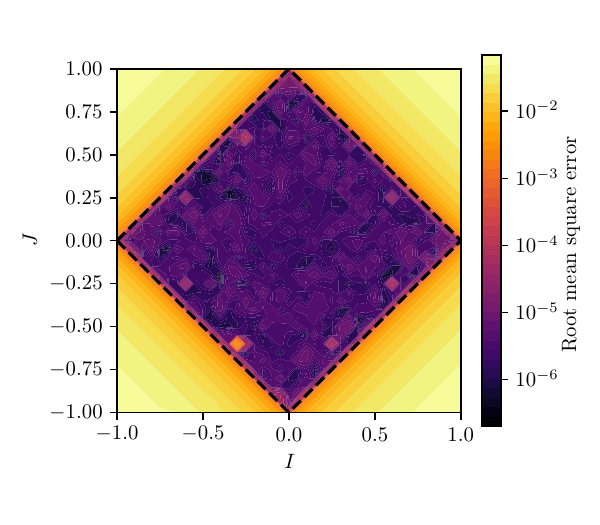}
\caption{Optimization error in the affine subspace spanned by the distributions (\ref{pX})--(\ref{p0 nontight}). The black dashed line is the boundary of the bilocal set in this two-dimensional slice of the behavior space, as proven in the end of this section. For behaviors inside the boundary, the solver can get the optimization error close to zero. As we move away from this region, the error grows, indicating that the behaviors are no longer bilocal.}
\label{fig:Bilocal error nontight}
\end{figure}

The BRGP inequality for this region of the behavior space reduces to $|X|\leq 1$. Once again, we will work in the domain $X,Y \in [-1,1]$, but in this case, all the points of the square satisfy the BRGP inequality, which leaves the whole square region as potentially bilocal. Let us check if this is indeed the case. The numerical procedure outlined above was performed for this affine subspace, with problem (\ref{optimization problem}) being solved 50 times for each behavior, with a density of 420 points per unit area of the $XY$ plane. The optimization error of this procedure can be seen in Fig. \ref{fig:Bilocal error nontight} (the black dashed line will be explained shortly).

We see in Fig. \ref{fig:Bilocal error nontight} that the solver found acceptable bilocal models only for points in the darker regions of the square domain, illustrating the non tightness of the BRGP inequality in this affine subspace. 
The square bounded by the black dashed lines in Fig. \ref{fig:Bilocal error nontight} is characterized by the following inequality:
\begin{equation}
|X| + |Y| \leq 1.\label{conjecture}
\end{equation}

Focusing on the boundary behaviors located in the first quadrant, their probability distribution is given by

\begin{multline}
p(a,b,c|x,y,z) = \frac{1}{8}\left(\frac{1}{2}+\delta_{a,0}\right)\\
\left[1 + X\delta_{y,0}(-1)^{a+b+c} + Y\delta_{y,1}(-1)^{z+a+b+c}\right],\label{boundary behaviors}
\end{multline}
with $X+Y = 1$. The resulting numeric models close to this line display the noticeable feature that two out of the four possible values of $\lambda$ have a probability very close to zero. This is an indication that $(c_\lambda,c_\mu)=(2,4)$ is enough to reproduce these probability distributions. By recalculating the models with these reduced cardinalities, and further noticing that all numerically determined entries are very close to simple rational numbers, we were able to devise the following model:
\begin{subequations}
\begin{align}
p_\lambda[\lambda] &= \begin{bmatrix}
    \frac{3}{4}& \frac{1}{4}
\end{bmatrix},\\
p_\mu[\mu] &= \begin{bmatrix}
    \frac{Y}{2} & \frac{X}{2} & \frac{Y}{2} & \frac{X}{2}
\end{bmatrix},\\
p_a[0,x,\lambda] &= \begin{bmatrix}
    1&0\\
    1&0
\end{bmatrix},\\
p_b[0,0,\lambda,\mu] &= \begin{bmatrix}
    \frac{1}{2}&1&\frac{1}{2}&0\\
    \frac{1}{2}&0&\frac{1}{2}&1
\end{bmatrix},\\
p_b[0,1,\lambda,\mu] &= \begin{bmatrix}
    1&\frac{1}{2}&0&\frac{1}{2}\\
    0&\frac{1}{2}&1&\frac{1}{2}
\end{bmatrix},\\
p_c[z,\mu] &= \begin{bmatrix}
    1&1&0&0\\
    0&1&1&0
\end{bmatrix}.
\end{align}
\end{subequations}
This model indeed reproduces (exactly) the behavior given by Eq. (\ref{boundary behaviors}). behaviors on the other three sides of the square ($X-Y=1$, $X+Y=-1$ and $X-Y=-1$) can be reproduced by the same model if we perform output and/or input relabellings. 

Therefore, we proved that all behaviors on the black dashed lines of Fig. \ref{fig:Bilocal error nontight} are bilocal. It turns out that the behavior at the center $X=Y=0$ is characterized by a factorized probability distribution
\begin{align*}
p_0(a,b,c|x,y,z) &= \left(\frac{1}{4} + \frac{1}{2}\delta_{a,0}\right)\left(\frac{1}{2}\right)\left(\frac{1}{2}\right)\nonumber\\
&= p(a|x)p(b|y)p(c|z).
\end{align*}

Thus, the projection of the bilocal set onto this affine subspace is star convex with respect to the center behavior $p_0$, as proven in Ref. \cite{bilocality_extended}. Therefore, it follows that all behaviors that satisfy inequality (\ref{conjecture}) are bilocal.

By applying inequality A2 of Ref. \cite{tripartite_bell_inequalities} to this affine subspace, we can see that inequality (\ref{conjecture}) is the boundary of the local set, where all parties have access to globally shared randomness. Therefore, since we found more restrictive bilocal models for the behaviors at this boundary it follows that inequality (\ref{conjecture}) is indeed the boundary of the bilocal set in this affine subspace.

The capability of identifying the boundary of the network-local set in these two cases serves as validation for the methodology we have been employing. Next we will show how the solutions of problem (\ref{optimization problem}) can lead to interesting results in the triangle scenario with no inputs.

\section{Triangle network}
\label{sec:triangle scenario numerical}

\begin{figure}
	\centering
\begin{tikzpicture}[scale=\scale,
		output/.style={circle, draw=green!60, fill=green!5, very thick, minimum size=10mm},
		hidden/.style={regular polygon, regular polygon sides=3, draw=red!60, fill=red!5, very thick, inner sep=0,minimum size=10mm},
		input/.style={rectangle,draw=blue!60, fill=blue!5, very thick, minimum size=6mm}
		]
		%Nodes
		\node[output] (a) at (0,0) {$a$};
		\node[hidden] (beta) at (60:2) {$\beta$};
		\node[hidden] (gamma) at (2,0) {$\gamma$};
		\node[output] (b) at (4,0) {$b$};
		\node[output] (c) at (60:4) {$c$};
		\path (60:4)+(-60:2) node[hidden] (alpha) {$\alpha$};
		
		%Edges
		\draw[-stealth] (beta) -- (a);
		\draw[-stealth] (beta) -- (c);
		\draw[-stealth] (gamma) -- (a);
		\draw[-stealth] (gamma) -- (b);
		\draw[-stealth] (alpha) -- (b);
		\draw[-stealth] (alpha) -- (c);
	\end{tikzpicture}
	\caption{Triangle scenario with no inputs. The outputs of each party $a$, $b$, and $c$ are represented by green circles. The local hidden variables $\alpha$, $\beta$, and $\gamma$ are represented by red triangles. Notice the lack of input choices for the parties. This is probably the simplest scenario that exhibits nonlocality without inputs.}
\label{fig:triangle scenario}
\end{figure}
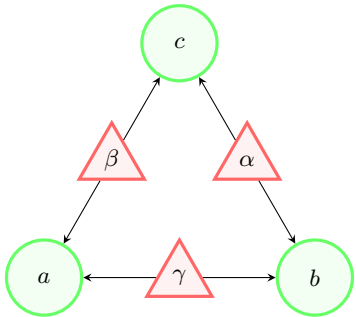

Consider the network scenario depicted in Fig. \ref{fig:triangle scenario}. This triangle scenario is the simplest network that can exhibit quantum nonlocality without any choice of inputs by the parties. Our goal in the following is to investigate specific regions of the behavior spaces of two and four output distributions. In Secs. \ref{subsec: GHZ} and \ref{subsec: W} we have $a,b,c\in\{0,1\}$, while in Sec. \ref{subsec: EJM} we have $a,b,c\in\{0,1,2,3\}$. For the triangle scenario with two possible outputs, the cardinality bound in the hidden variables is $6$, whereas for the case with four possible outputs it is $60$ \cite{cardinality_bound}.

\subsection{GHZ distribution}
\label{subsec: GHZ}

First, let us consider the GHZ distribution, named after the homonymous quantum state \cite{GHZ}, mixed with a uniformly distributed random noise:

\begin{equation}
p(a,b,c) = \begin{cases}\frac{v}{2} + \frac{1-v}{8}&\text{if }a=b=c\\
\frac{1-v}{8}&\text{otherwise.}\end{cases}\label{GHZ distribution} 
\end{equation}

\begin{figure}
	\centering
\includegraphics[width=\columnwidth]{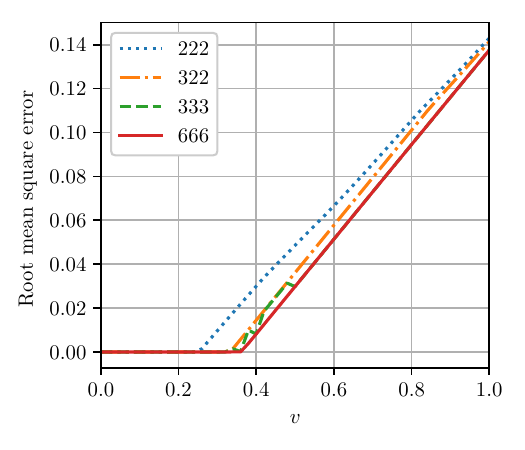}
\caption{Optimization error as a function of the visibility of the GHZ distribution. For visibilities $v\lessapprox0.36$, the error reaches values very close to zero, indicating that these behaviors are network local. For visibilities $v\gtrapprox0.36$, the solver can no longer achieve low optimization errors, indicating that these behaviors are most likely not network local.}
\label{fig:GHZ error}
\end{figure}

We solved problem (\ref{optimization problem}) for this distribution with the cardinalities $(c_\alpha,c_\beta,c_\gamma)=(2,2,2)$, $(3,2,2)$, $(3,3,3)$, and $(6,6,6)$, the latter corresponding to the maximal necessary cardinality. The optimization error as a function of the visibility for each case can be seen in Fig. \ref{fig:GHZ error}. The critical visibility increases as we allow for more complex models, as expected, but only up to $(3,3,3)$ models. Notice that the results for the triplet $(3,3,3)$ typically agree with those of the triplet $(6,6,6)$, with some oscillations to the $(3,2,2)$ curve. Obviously, this is a numeric artifact that could be eliminated by increasing the number of trials per behavior.

If we inspect the models for regions close to the critical visibility in each case, several entries in the response function arrays are very close to 0 or 1. By fixing those entries and solving for the other ones with the use of a symbolic computation package, we were able to find the following models for each cardinality triplet.

For $(c_\alpha,c_\beta,c_\gamma)=(2,2,2)$, the following model reproduces the GHZ distribution with visibility $v=1/4$:
\begin{subequations}
\begin{align}
    &p_\alpha[\alpha] = p_\beta[\beta] = p_\gamma[\gamma] = \begin{bmatrix}\frac{1}{2} & \frac{1}{2}\end{bmatrix}\label{hidden variables GHZ 222 crit}\\
    &p_a[0,\beta,\gamma] = p_b[0,\gamma, \alpha] = p_c[0,\alpha, \beta] = \begin{bmatrix}0 & \frac{1}{2}\\
    \frac{1}{2} & 1\end{bmatrix}\label{response functions GHZ 222 crit}
\end{align}
\end{subequations}
Lower visibilities can be achieved if the parties follow this strategy with some probability or just produce uniformly distributed random outputs, otherwise.

For $(c_\alpha,c_\beta,c_\gamma)=(3,2,2)$, the following model reproduces the target distribution for all visibilities in the range $0\leq v\leq 1/3$:
\begin{subequations}
\begin{align}
    p_\alpha[\alpha] &= \begin{bmatrix}\frac{v}{2} & 1-v & \frac{v}{2}\end{bmatrix},\\
    p_\beta[\beta] &= \begin{bmatrix}\frac{1}{2} & \frac{1}{2}\end{bmatrix},\\
    p_\gamma[\gamma] &= \begin{bmatrix}\frac{1}{2} & \frac{1}{2}\end{bmatrix},\\
    p_a[0,\beta,\gamma] &= \begin{bmatrix}\frac{1-3v}{2(1-v)} & \frac{1}{2}\\
    \frac{1}{2} & \frac{1+v}{2(1-v)}\end{bmatrix},\\
    p_b[0,\gamma, \alpha] &= \begin{bmatrix}1& 0 & 0\\
    1& 1 & 0\end{bmatrix},\label{Bob response GHZ}\\
    p_c[0,\alpha, \beta] &= \begin{bmatrix}1 & 1\\
    0 & 1\\
    0 & 0\end{bmatrix}.\label{Charles response GHZ}
\end{align}
\end{subequations}

For $(c_\alpha,c_\beta,c_\gamma) = (3,3,3)$, we found the following model with parameters $a$ and $b$:
\begin{subequations}
\begin{align}
    &p_\alpha[\alpha] = p_\beta[\beta] = p_\gamma[\gamma] = \begin{bmatrix}a & b & 1-a-b\end{bmatrix},\label{hidden variables GHZ 333 crit}\\
    &p_a[0,\beta,\gamma] = p_b[0,\gamma, \alpha] = p_c[0,\alpha, \beta] = \begin{bmatrix}0 & 1 & 0\label{response functions GHZ 333 crit}\\
    1 & 1 & 0 \\
    0 & 0 & 0\end{bmatrix}.
\end{align}
\end{subequations}
By setting $a\approx 0.454$ and $b\approx 0.386$, the unique roots in the interval $[0,1]$ of the polynomials $12a^4-8a^3+6a^2-1$ and $4b^4-8b+3$, respectively, the model reproduces the behavior of Eq. (\ref{GHZ distribution}) with visibility 
$$v = 2b-3+\frac{1}{b} \approx 0.362.$$ 
This visibility is the largest real root of the polynomial $3v^4 + 28 v^3 + 66v^2 - 36v +3$. Since the model is symmetric with respect to the parties, lower visibilities can be achieved in the same manner of the $(2,2,2)$ models, that is, with appropriate white noise admixture. 

These findings lead us to conjecture that the critical visibilities for GHZ distribution in the context of each cardinality triplet are $v_c^{222}=1/4$, $v_c^{322}=1/3$, and $v_c^{333}\approx0.362$, matching our expectations from Fig. \ref{fig:GHZ error}. We also conjecture that no further increases in the critical visibility are possible by considering models with cardinalities higher than $(3,3,3)$, as Fig. \ref{fig:GHZ error} suggests.

It is important to stress the fact that although the GHZ distribution is named after the well-known quantum state, this does not mean that it is realizable with visibility $v=1$ in the triangle scenario using quantum sources, because the GHZ state exhibits tripartite entanglement, while we are only allowed to use bipartite sources in the triangle network. Indeed, using inequality (34) of Ref. \cite{inflation}, it can be proven that even using quantum sources, $v\leq 0,5$. This indicates that it may be possible for noisy GHZ distributions of quantum mechanical origin to be network nonlocal (for $0.362<v\le 0.5$).

What we did in the above was to try to fit a network-local model to a distribution $p=vp_1 + (1-v)p_0$, where $p_1$ is the GHZ distribution and $p_0$ is white noise. If the distributions $p_1$ and $p_0$ exhibit symmetry between the three parties (as is the case), and the network-local set is not too much concave, we expect the local model for $v=v_c$ to be the solution found by the optimization routine even when $v>v_c$. In this situation, it is impossible to find a local model, but the model that best approximates the target distribution is the one for $v=v_c$. This is expected at least in close proximity of $v_c$. If that is the case, the root-mean-square error in the probabilities is:

\begin{align}
E &= \left\{\frac{1}{8}\sum_{a,b,c}\left[p_c(a,b,c) - p^*(a,b,c)\right]^2\right\}^{1/2}\nonumber\\
&= \left\{\frac{1}{8}\sum_{a,b,c}\left[p_0(a,b,c) - p_1(a,b,c)\right]^2\right\}^{1/2}(v-v_c).\label{error slope}
\end{align}

We see that the root-mean-square error should increase linearly with $v$, as shown in Fig. \ref{fig:GHZ error}. Indeed, for the GHZ distribution, Eq. (\ref{error slope}) reduces to 
$$E = \frac{\sqrt{3}}{8}(v-v_c) \approx 0.2165(v-v_c),$$ 
which matches the slope seen in Fig. \ref{fig:GHZ error} for models with $(c_\alpha,c_\beta,c_\gamma) = (6,6,6)$ (also for models with $c_\alpha=c_\beta=c_\gamma\ge 3$).

\subsection{W distribution}
\label{subsec: W}

Let us consider another target behavior, the so called W distribution, named after the homonymous quantum state \cite{W}, mixed with uniformly distributed random noise:

\begin{equation}
p(a,b,c) = \begin{cases}\frac{v}{3} + \frac{1-v}{8}&\text{if }a+b+c=1\\
\frac{1-v}{8}&\text{otherwise.}\end{cases}    
\end{equation}

\begin{figure}
	\centering
\includegraphics[width=\columnwidth]{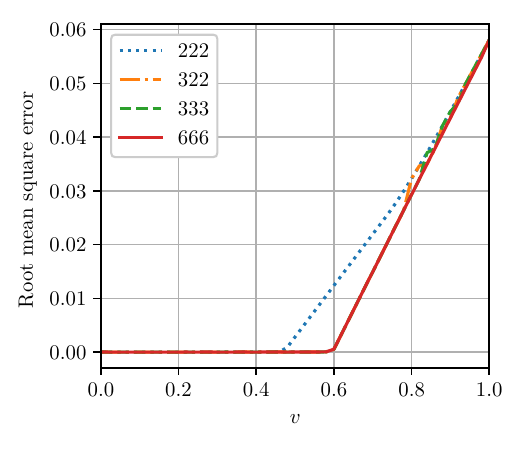}
\caption{Optimization error as a function of the visibility of the W distribution. For visibilities $v\lessapprox 0.6$, the error reaches values very close to zero, indicating that these behaviors are network local. For visibilities $v \gtrapprox 0.6$, the solver can no longer achieve low optimization errors, indicating that these behaviors are most likely not network local.}
\label{fig:W error}
\end{figure}

The root-mean-square error in the probabilities can be seen in Fig. \ref{fig:W error}, for the same cardinatilities as in Sec. \ref{subsec: GHZ}. 
Unlike what happened with the GHZ distribution, the cardinalities $(c_\alpha, c_\beta, c_\gamma) = (3,2,2)$ seem to be enough to solve the problem. Again, we see some oscillations of the $(3,2,2)$ curve between those of the triplets $(2,2,2)$ and $(6,6,6)$. By running the optimization for $v$ close to the critical value, we obtain a model with $\beta$ and $\gamma$ being nearly identically distributed and Bob and Charles having nearly deterministic response functions. By forcing these constraints and solving for the other parameters, we were able to find the following model for $v \in [0,v_c]$:

\begin{subequations}
\begin{align}
    p_\alpha[\alpha] &= \begin{bmatrix}\frac{3+v}{12} - \frac{1-v}{4u} & \left(\frac{1-v}{4}\right)\left(1 + \frac{1}{u}\right)^2& \frac{3+v}{12} - \frac{1-v}{4u}\end{bmatrix}\\
    p_\beta[\beta] &= p_\gamma[\gamma] = \begin{bmatrix}\frac{u}{1+u} & \frac{1}{1+u}\end{bmatrix}\\
    %p_\gamma[\gamma] &= \begin{bmatrix}\frac{u}{1+u} & \frac{1}{1+u}\end{bmatrix}\\
    p_a[0,\beta,\gamma] &= \begin{bmatrix}\frac{1}{2} & \frac{3+v}{6} + \frac{uv(9-v)}{18(1-v)}\\
    \frac{3+v}{6} + \frac{uv(9-v)}{18(1-v)} & \frac{3(1-v)}{2(3+v)}\end{bmatrix}\\
    p_b[0,\gamma, \alpha] &= \begin{bmatrix}1& 0 & 0\\
    1& 1 & 0\end{bmatrix}\label{Bob response W}\\
    p_c[0,\alpha, \beta] &= \begin{bmatrix}0 & 0\\
    0 & 1\\
    1 & 1\end{bmatrix},\label{Charles response W}
\end{align}
\end{subequations}
where $u = \sqrt{3(1-v)/(3+v)}$. This model breaks down for $v>v_c$ because the off-diagonal elements of Alice's response function become larger than 1. By constraining these elements to be equal to 1, one finds that $v_c$ is the root of the polynomial $4v^4 - 30v^3 + 63v^2 + 108v - 81$ which is within the interval $[0,1]$, i.e., $v_c \approx 0.5966$, matching our expectations from Fig. \ref{fig:W error}.

As was the case of the GHZ distribution, notice that the W distribution with $v=1$ is not achievable using quantum bipartite sources in the triangle scenario. Indeed, it has been proven that quantum strategies cannot achieve visibilities higher than $v= 6-3\sqrt{3}\approx0.8039$ \cite{quantum_inflation}. These results also leave room for quantum-mechanical noisy W distributions to present non-trilocality.

The slope of Fig. \ref{fig:W error} for the region $v>v_c$ can be analyzed in the same way as we did in Sec. \ref{subsec: GHZ}. For the W distribution, Eq. (\ref{error slope}) becomes $$E = \frac{\sqrt{15}}{24}(v-v_c) \approx 0.1614(v-v_c).$$ This slope matches the one obtained numerically in Fig. \ref{fig:W error} for models with $(c_\alpha,c_\beta,c_\gamma)=(6,6,6)$ (also for models with $c_\alpha=c_\beta+1=c_\gamma+1\ge 3$).

One interesting difference between the GHZ and W distributions is that the model for the W distribution is not symmetric, while the one for the GHZ distribution is. This has prompted us to run optimizations for the W distribution subject to the constraint that the model should be symmetric. With this additional requirement, the cardinality bound of Eq. (\ref{cardinality bound}) is no longer valid, but even for cadinalities as high as $c_\alpha=c_\beta=c_\gamma=23$, we have only been able to reach $v\approx0.51$. Also, this maximum visibility remains fairly stable after $c_\alpha = c_\beta = c_\gamma = 4$, an indication that it might not be possible to reach $v= v_c \approx 0.5966$ with symmetric models. At the moment, we are unsure whether there is a deeper reason for this difference in the local models for the GHZ and W distributions, despite the fact that in both cases the final probability distribution is completely symmetric with respect to the three parties.

\subsection{EJM distribution}
\label{subsec: EJM}
Finally, we consider the so called elegant joint measurement (EJM) distribution, that can be generated by a quantum strategy \cite{elegant_joint_measurement}. This behavior is conjectured to be nonlocal, so a natural question is: how much noise must we mix with the EJM distribution so that it becomes network local? The target distribution is:

\begin{equation}
    p(a,b,c) = vp_{EJM}(a,b,c) + \frac{1-v}{64}.\label{EJM distribution}
\end{equation}

For the triangle network with four outputs, the full cardinalities $c_\alpha = c_\beta = c_\gamma = 60$ render the optimization run time prohibitively large. However, some insight into the critical visibility for a given distribution can be gained by exploring models with lower cardinalities. We have tested for all possible cardinalities up until $c_\alpha = c_\beta = c_\gamma = 7$. The tables below show the maximum values of $v$ for which the numerical optimization procedure could find a model with root-mean-square error less than $1\times10^{-4}$. We believe these should serve as rough estimates for the critical visibility for network-local behaviors constrained by the tested cardinalities. Notice that the target distribution is symmetric with respect to the three parties, so we can without loss of generality assume henceforth that $c_\alpha \geq c_\beta \geq c_\gamma$ 
(that justifies the blank spaces in the tables).

\bigskip
\noindent
$c_{\gamma}=2$\\
\begin{tabular}{|c||*{6}{c|}}\hline
\backslashbox{$c_\beta$}{$c_\alpha$}
&\makebox[3em]{2}&\makebox[3em]{3}&\makebox[3em]{4}
&\makebox[3em]{5}&\makebox[3em]{6}&\makebox[3em]{7} \\\hline\hline
2 & 0.005 & 0.007 & 0.009 & 0.012 & 0.012 & 0.012\\\hline
3 &       & 0.012 & 0.014 & 0.014 & 0.014 & 0.014\\\hline
4 &       &       & 0.018 & 0.018 & 0.018 & 0.018 \\\hline
5 &       &       &       & 0.018 & 0.018 & 0.018 \\\hline
6 &       &       &       &       & 0.018 & 0.018 \\\hline
7 &       &       &       &       &       & 0.018 \\\hline
\end{tabular}

\bigskip

\bigskip
\noindent
$c_{\gamma}=3$\\
\begin{tabular}{|c||*{5}{c|}}\hline
\backslashbox{$c_\beta$}{$c_\alpha$}
&\makebox[3em]{3}&\makebox[3em]{4}
&\makebox[3em]{5}&\makebox[3em]{6}&\makebox[3em]{7} \\\hline\hline
3        & 0.014 & 0.018 & 0.018 & 0.018 & 0.018\\\hline
4        &       & 0.025 & 0.025 & 0.025 & 0.025 \\\hline
5        &       &       & 0.025 & 0.025 & 0.025 \\\hline
6        &       &       &       & 0.025 & 0.025 \\\hline
7        &       &       &       &       & 0.025 \\\hline
\end{tabular}

\bigskip

\bigskip
\noindent
$c_{\gamma}=4$\\
\begin{tabular}{|c||*{4}{c|}}\hline
\backslashbox{$c_\beta$}{$c_\alpha$}
&\makebox[3em]{4}
&\makebox[3em]{5}&\makebox[3em]{6}&\makebox[3em]{7} \\\hline\hline
4 & 0.318 & 0.397 & 0.397 & 0.397 \\\hline
5 &       & 0.398 & 0.399 & 0.399 \\\hline
6 &       &       & 0.399 & 0.400 \\\hline
7 &       &       &       & 0.400 \\\hline
\end{tabular}

\bigskip

\bigskip
\noindent
$c_{\gamma}=5$\\
\begin{tabular}{|c||*{3}{c|}}\hline
\backslashbox{$c_\beta$}{$c_\alpha$}
&\makebox[3em]{5}&\makebox[3em]{6}&\makebox[3em]{7} \\\hline\hline
5 & 0.398 & 0.402 & 0.403 \\\hline
6 &       & 0.402 & 0.405 \\\hline
7 &       &       & 0.405 \\\hline
\end{tabular}

\bigskip

\bigskip
\noindent
$c_{\gamma}=6$\\
\begin{tabular}{|c||*{2}{c|}}\hline
\backslashbox{$c_\beta$}{$c_\alpha$}
&\makebox[3em]{6}&\makebox[3em]{7} \\\hline\hline
6 & 0.404 & 0.405 \\\hline
7 &       & 0.405 \\\hline
\end{tabular}

\bigskip

\bigskip
\noindent
$c_{\gamma}=7$\\
\begin{tabular}{|c||*{1}{c|}}\hline
\backslashbox{$c_\beta$}{$c_\alpha$}
&\makebox[3em]{7} \\\hline\hline
7 & 0.405 \\\hline
\end{tabular}
\bigskip

As can be noticed by inspecting the tables, the smallest cardinality $c_\gamma$ seems to be the most constraining factor to the maximum visibility that one can achieve using network-local models. Furthermore, we see a significant increase in the critical visibility from values close to zero to $0.318$ for cardinalities $(c_\alpha, c_\beta, c_\gamma) = (4,4,4)$. After that, there is another increase in the critical visibility to $0.397$ for $(c_\alpha, c_\beta, c_\gamma) = (5,4,4)$, and then the increments become fairly marginal. These results show that for visibilities up to approximately 0.4, the EJM distribution mixed with uniform random noise is trilocal. It is always possible that this range can be widened by utilizing larger cardinalities, but the approximate value of 0.4 is a lower bound for the true critical visibility. It can be regarded as an initial candidate to the true critical visibility, regardless of the hidden-variable cardinalities.

\begin{figure}[h]
	\centering
	\includegraphics[width=\columnwidth]{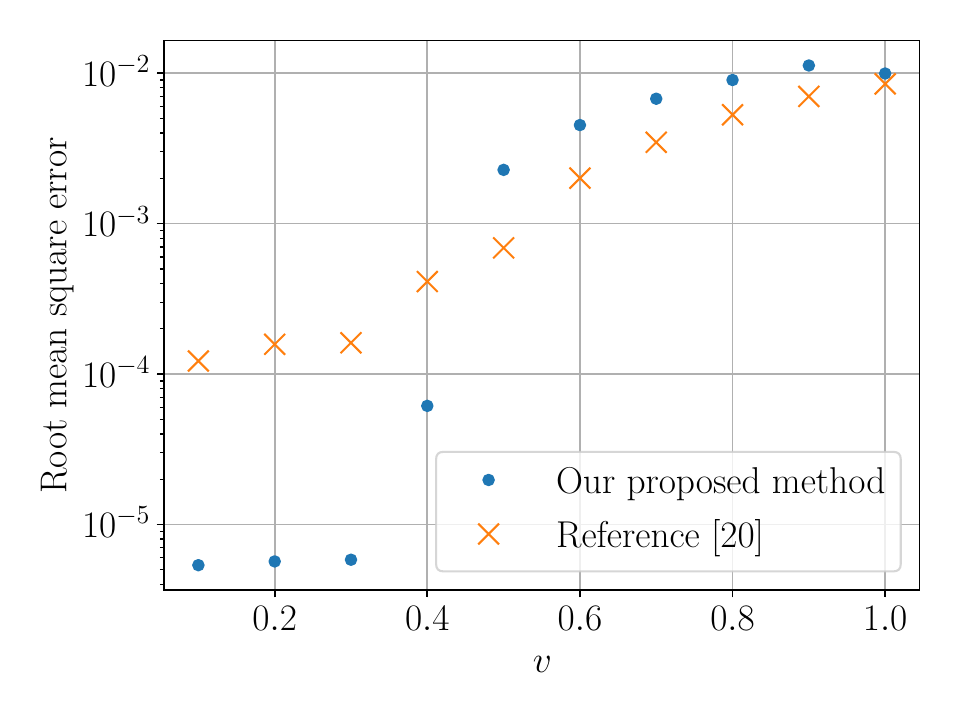}
	\caption{Comparison of the optimization errors of our method (with cardinalities $c_\alpha=c_\beta=c_\gamma=7$) and the one of Ref. \cite{neural_network}.}\label{fig: comparison neural networks}
\end{figure}

We had mentioned before that this work is in the same spirit of the neural network optimization put forth in Ref. \cite{neural_network}. Since the authors also discuss the EJM distribution, albeit with different noise models than the one considered here, a direct comparison between the two methodologies is in order. Both optimization procedures were run with visibilities $v = 0.1,0.2,\ldots, 1$, and the code used in \cite{neural_network} was slightly altered to target distributions with our model of global noise, as in Eq. (\ref{EJM distribution}). For our method, we have chosen to optimize over models with cardinalities $c_\alpha=c_\beta=c_\gamma=7$, and with 20 trials per behavior. With these settings, both programs took about the same time to finish using the same hardware (about 7 hours in an average personal computer), and the results can be seen in Fig. \ref{fig: comparison neural networks}.

It is interesting that both optimization strategies show an increase in the root-mean-square error when the visibilities get near $v=0.4$, which is more evidence that the true critical visibility for the EJM distribution is close to this mark. One other aspect that we would like to point out is that without tweaking the optimization parameters and just running the default versions of the two pieces of software as we did in this test, our method gave lower errors within the local set but higher ones in the nonlocal region. This indicates that the optimization routine we propose can yield a better contrast between the local and nonlocal portions of the behavior space.

\section{Closing remarks}
\label{closingR}

A tool for finding explicit models in network scenarios with finite hidden variable cardinality was developed. We hope it will be useful to the community interested in advancing the understanding of network-nonlocality. In this paper, we have used the tool to find the critical visibilities of the GHZ, W, and EJM distributions, when mixed with uniform random noise. For the GHZ and W distributions, we found fourth degree polynomials whose roots are conjectured to be the critical visibilities, along with the exact network-local strategies that generate these behaviors. For the EJM distribution, we found only numerical estimates of the critical visibility by looking at models with reduced cardinalities for the hidden variables.

The first area where future work could be done is in trying to improve the execution time and accuracy of the numerical routine. In the implementation used to generate Figs. \ref{fig:Bilocal error}, \ref{fig:Bilocal error nontight}, \ref{fig:GHZ error} and \ref{fig:W error}, each time the solver attempts to find a local model, the initial guess is completely random. So if an acceptable model was found for one point, this achievement was not taken into account when  investigating neighboring points. Doing this could help expedite calculations and also reduce the probability of failure when trying to find a model for a point that is network-local. In certain affine subspaces, symmetry could also be an important factor. For example, in the subspace of Fig. \ref{fig:Bilocal error}, the symmetries $I\rightarrow -I$ and $J \rightarrow -J$ correspond to simple relabelings of outputs and inputs. So there is actually no need to employ computing power in more than one quadrant.

An interesting issue is the matter of cardinality reduction in the hidden variables. As we saw in Sec. \ref{sec:triangle scenario numerical}, the behavior obtained by the GHZ distribution mixed with the uniform distribution with critical visibility could be obtained with cardinalities $c_\alpha=c_\beta=c_\gamma=3$, although the upper bound provided by \cite{cardinality_bound} is $c_\alpha=c_\beta=c_\gamma=6$. On the other hand, for the W distribution, $(c_\alpha,c_\beta,c_\gamma) = (3,2,2)$ seems to be sufficient to reproduce the critical behavior. What determines when such reductions are possible and by how much? An investigation of this question could shed light on network nonlocality.

The fact that the critical visibilities of both the GHZ and W distributions are conjectured to be roots of 4th degree polynomials tells us that there would be polynomial Bell inequalities of the same degree (at least) for the triangle scenario with binary outputs that still eludes us. The search for these inequalities is also an exciting possible avenue of future research.

\begin{acknowledgments}
This work received financial support from the Brazilian agencies Coordena\c{c}\~ao de Aperfei\c{c}oamento de Pessoal de N\'{\i}vel Superior (CAPES), Funda\c{c}\~ao de Amparo \`a Ci\^encia e Tecnologia do Estado de Pernambuco (FACEPE), Conselho Nacional de Desenvolvimento Cient\'{\i}fico  e Tecnol\'ogico through its program CNPq INCT-IQ (Grant 465469/2014-0), and Funda\c{c}\~ao de Amparo \`a Pesquisa do Estado de S\~ao Paulo (FAPESP - Grant 2021/06535-0).
\end{acknowledgments}

\bibliography{references2.bib}

%%%%%%%%%%%%%%%%%%%%%%%%%%%%%%%%%%%%%%%%%%%%%%%%%%%%%%%%%%%%

\end{document}